\documentclass[usenatbib,conference]{basi}
\usepackage[T1]{fontenc}
\usepackage[british]{babel}
\usepackage[varg]{txfonts}
\usepackage{rotating}
\usepackage{dcolumn}
\begin{document}
\title[Spectral and morphological signatures of cluster merger shocks]{Spectral and morphological signatures of cluster merger shocks: CIZA J2242.8+5301}
\author[A.~Stroe et~al.]%
       {A.~Stroe$^1$\thanks{email: \texttt{astroe@strw.leidenuniv.nl}},
       R.~van~Weeren$^{2}$, D.~Sobral$^{1,3}$, C.~Rumsey$^{4}$, H.~Intema$^{5}$,
       \newauthor
        H. ~R\"ottgering$^{1}$, J.~Harwood$^{6}$, R.~Saunders$^{4,7}$, M.~Br\"uggen$^{8}$ and M.~Hoeft$^{9}$\\
       $^{1}$Leiden Observatory, Leiden University, P.O.\ Box 9513, NL-2300 RA Leiden, The Netherlands\\
       $^{2}$Harvard Smithsonian Center for Astrophysics (CfA - SAO), 60 Garden Street Cambridge, MA 02138\\
       $^{3}$CAAUL, Observat\'{o}rio Astron\'{o}mico de Lisboa, Tapada da Ajuda, 1359-018, Lisboa, Portugal\\ 
       $^{4}$Astrophysics Group, Cavendish Laboratory, JJ Thomson Avenue, Cambridge, CB3 0HE\\
       $^{5}$NRAO, Pete V. Domenici Science Operations Center, 1003 Lopezville Road, Socorro, NM 87801-0387, USA\\
       $^{6}$University of Hertfordshire, College Lane, Hatfield, Hertfordshire AL10 9AB, UK\\
       $^{7}$Kavli Institute for Cosmology Cambridge, Madingley Road, Cambridge CB3 0HA, UK\\
       $^{8}$Hamburger Sternwarte, Gojenbergsweg 112, 21029 Hamburg, Germany\\
       $^{9}$Th\"uringer Landessternwarte Tautenburg, Sternwarte 5, 07778, Tautenburg, Germany}
\pubyear{2014}
\volume{00}
\pagerange{\pageref{firstpage}--\pageref{lastpage}}
\status{submitted}
\maketitle

\label{firstpage}
   
\begin{abstract}
The CIZA J2242.8+5301 cluster hosts an extraordinary, narrow, Mpc-wide, diffuse patch of radio emission tracing travelling shock waves, called a relic. We perform a multi-wavelength of observations of the cluster. We discover radio spectral steepening and increasing spectral curvature in the shock downstream area, indicative of emission produced by spectrally-aged electrons, following a JP ageing model. The curved integrated spectrum towards $16$ GHz challenges the favoured relic formation model. Further, we find a boosting the number density of H$\alpha$ emitters by a factor of $10$, possibly caused by interactions of the galaxies with the shock wave.
\end{abstract}
\section{Introduction}\label{s:intro}
\begin{figure}
\begin{center}
\begin{tabular}{p{7cm}cp{4cm}}
\raisebox{-\height}{\includegraphics[width=7.0cm]{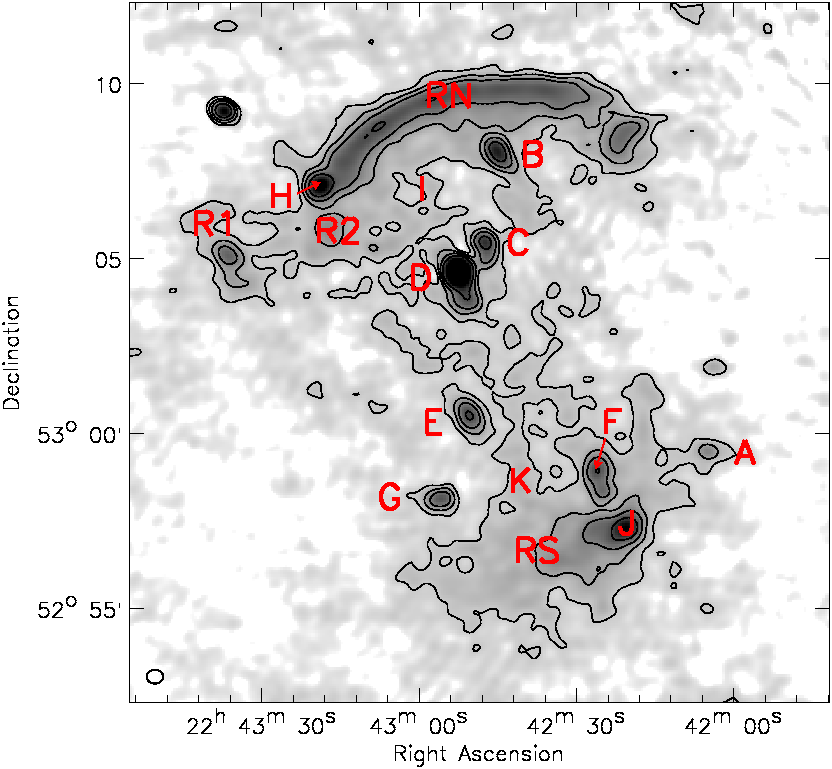}} & \quad &
\caption{$153$~MHz image of the cluster with contours drawn at ${[4,8,16,32]} \times \sigma_{\mathrm{RMS}}$ ($\sigma_{\mathrm{RMS}}=1.5$ mJy) and source labelling. Beam shown in the bottom-left corner \citep[image from ][]{2013A&A...555A.110S}. \label{fig:radio}}
\end{tabular}
\end{center}
\end{figure}

Giant radio relics are diffuse, elongated, Mpc-wide, polarised patches of emission found at the outskirts of galaxy clusters \citep[e.g. review by][]{2012A&ARv..20...54F}. During major cluster mergers, travelling shock waves are thought to form, which would (re-)accelerate intra-cluster medium (ICM) particles to relativistic speeds via the diffusive-shock acceleration mechanism \citep[DSA; e.g.][]{1983RPPh...46..973D, 1998A&A...332..395E}. In the presence of $\mu$G-level magnetic fields, these electrons emit synchrotron radiation and, with time, lose energy via synchrotron and inverse Compton processes. One of the best-studied examples of clusters with radio relics is CIZA J2242.8+5301 (nicknamed the `sausage') at $z=0.192$ \citep{2010Sci...330..347V}. Towards the north of the cluster, a thin ($50$ kpc), regular, $1.4$-Mpc long radio relic is found (see Fig.~\ref{fig:radio}, labelled RN). \citet{2013MNRAS.429.2617O} also found evidence for a shock from in the X-rays at the same location as the radio. 
The cluster also hosts a variety of diffuse radio sources (e.g. R1, R2, RS, J in Fig~\ref{fig:radio}) and tailed radio galaxies \citep[labelled B through F;][]{2013A&A...555A.110S}. We perform a broad-band radio analysis, spanning from $153$~MHz to $16$~GHz, to  uncover the injection and ageing mechanisms involved in the production of the `sausage' relic. We also investigate the interaction between the travelling shock front and the galaxies within the cluster, to test whether the wave inhibits or triggers star-formation.

\section{Shock influence on the ICM - radio relic spectral properties}
To trace the northern relic properties, we have observed the cluster with the Giant Metrewave Radio Telescope (GMRT), the Westerbork Synthesis Radio Telescope (WSRT) and the Arcminute Microkelvin Imager (AMI) at eight, almost logarithmically-spaced frequencies between $153$ MHz and $16$ GHz. The $153$ MHz map with source labelling can be found in Fig.~\ref{fig:radio}. By combining $18''\times14''$ resolution GMRT and WSRT data, we produce RN spectral index $\alpha$ and spectral curvature $C$ maps and find clear signs of steepening and increasing curvature from the shock location at the north of the relic, towards the south, into the downstream area. These findings strongly support a DSA-like scenario, where electrons are injected on one side, likely a shock, and subsequently age. For regions within the northern relic selected on their similar $\alpha$ properties, we build a colour-colour plot, in which their low and high frequency spectral indeces are plotted against each other \citep{1993ApJ...407..549K}. This allows comparison with predictions of spectral ageing models. The best fit is a superposition of Jaffe-Perola models \citep[JP;][]{1973A&A....26..423J}: a single burst of electron injection is followed by cooling with electron pitch angle isotropisation in a uniform magnetic field. Because of projection effects, each image pixel consists of a range of electron populations with different ages \citep[time since last acceleration;][]{2013A&A...555A.110S}.

The integrated RN spectrum ($153$ MHz to $2.3$ GHz) is well described by a single power law of index $-1.06\pm0.04$ (see left panel of Fig.~\ref{fig:2}). However, towards $16$ GHz, the spectrum curves, measuring $12\sigma$ below the prediction from simple `stationary conditions' \citep[i.e. continuous injection, CI, spectral model;][]{pacholcyzk}, which assumes that the relic has been in the ICM longer than the electron cooling time. This could be caused by a non-power-law injection spectrum, affected by radiation losses, or an ICM inhomogeneity: a density and/or temperature gradient across the relic would lead to a steady modification of the Mach number or to a decrease in levels of injected electrons with time. Highly-turbulent magnetic fields in the downstream area would have a similar effect \citep{AMI}.

\section{Shock influence on galaxies - H$\alpha$ luminosity function}
\begin{figure}
\centerline{\includegraphics[width=0.53\textwidth]{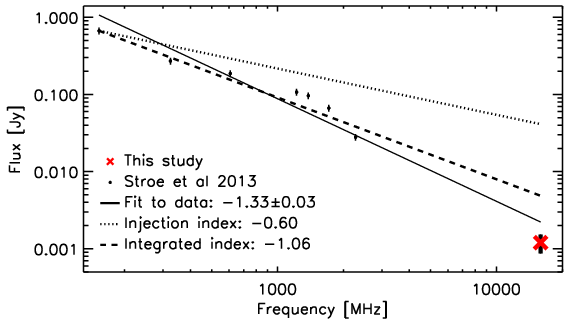} \qquad
\includegraphics[width=0.41\textwidth]{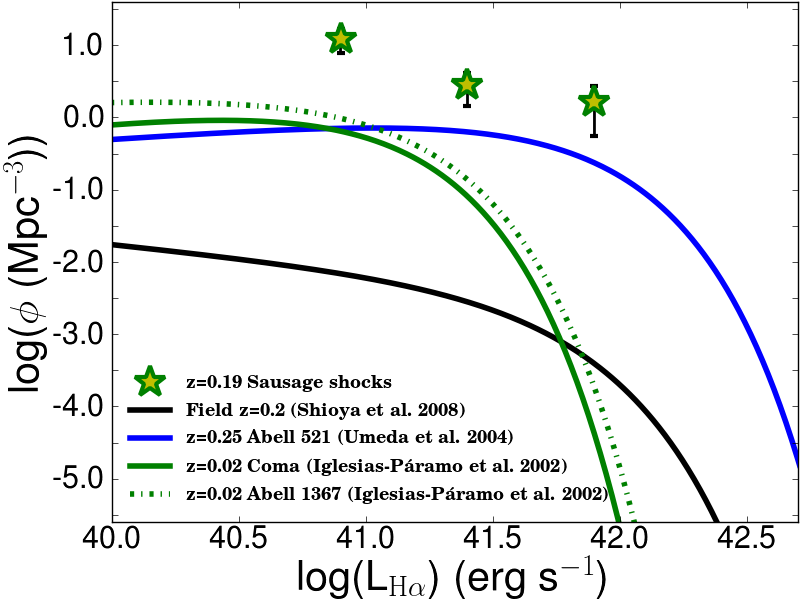}}
\caption{Left: Integrated radio spectrum of RN from $153$~MHz up to $16$~GHz. The $16$ GHz measurement is $12\sigma$ away from the DSA prediction. Right: H$\alpha$ narrow-band LF for the `sausage' cluster, calculated for the regions projected around the relics. We overlay the LFs for the field at $z=0.2$ and clusters Abell 521 at $0.25$, Abell 1367 and Coma at $z=0.02$.
\label{fig:2}}
\end{figure}
To investigate the effect of the shock on the cluster galaxies, we observe the H$\alpha$ emission line, a well-calibrated star-formation tracer. Using a custom-made narrow-band optical filter mounted on the Isaac Newton telescope, we survey the H$\alpha$ emission at the cluster redshift. We find $\sim15$ galaxies with extended (up to $60$~kpc), disturbed H$\alpha$ haloes marked by tails and bright star-forming regions in close proximity of the diffuse sources which trace shock waves. To quantify this effect, we build an H$\alpha$ luminosity function (LF, see Fig.~\ref{fig:2}, right panel). The H$\alpha$ counts in the range of the `sausage' relic areas are an order of magnitude ($9\sigma$ significance) above the LFs of other clusters. For low-redshift, relaxed clusters Abell 1367 and Coma \citep{2002A&A...384..383I}, the characteristic luminosity is below the blank field measurement in the COSMOS field \citep{2008ApJS..175..128S}, indicating a lack of bright emitters. However, the merging cluster Abell 521 ($z=0.25$) contains many luminous H$\alpha$ galaxies \citep{2004ApJ...601..805U}. We speculate that the enhancement in H$\alpha$ emission within the cluster galaxies around the relics is related to the merging activity, via shock-induced star-formation \citep{1989MNRAS.239P...1R}. The passage of the shock front might temporarily compress the low-temperature gas, which then collapse into star-forming clouds \citep{2014MNRAS.438.1377S}.

\section{Conclusion}
We performed a radio-optical analysis of the `sausage' cluster, which hosts a regular, arc-shaped radio relic. We find that the spectral curvature of the radio emission systematically increases from the shock location into the downstream area. This effect is predicted by the preferred relic formation mechanism, in which the travelling shock accelerates ICM electrons via DSA. Nevertheless, $16$ GHz data indicates steepening of the integrated relic spectrum, which is in stringent tension with DSA. The results could be reconciled by ICM temperature/density gradients across the source or a turbulent downstream magnetic field. An analysis of the cluster H$\alpha$ galaxies indicates that the shock front had an important role in their evolution, possibly compressing the gas, which started forming stars. Therefore, travelling shock fronts created at cluster mergers not only interact with the ICM, leading to the formation of giant radio relics, but may also shape the properties of cluster galaxies. 

\bibliography{Proceeding_Stroe_arxiv}

\begin{thebibliography}{}
\bibitem[{{Drury}(1983)}]{1983RPPh...46..973D}
{Drury}, {L.~O.} 1983, Reports on Progress in Physics, 46, 973

\bibitem[{Ensslin} {et~al.}(1998)]{1998A&A...332..395E}
{Ensslin}, {T.~A.}, {Biermann}, {P.~L.}, {Klein}, {U.}, \& {Kohle}, {S.} 1998, A\&A, 332, 395

\bibitem[{Feretti} {et~al.}(2012)]{2012A&ARv..20...54F}
{Feretti}, {L.}, {Giovannini}, {G.}, {Govoni}, {F.}, \& {Murgia}, {M.} 2012, A\&Ar, 20, 54

\bibitem[Iglesias-P{\'a}ramo et al.(2002)]{2002A&A...384..383I} 
Iglesias-P{\'a}ramo, J., Boselli, A., Cortese, L., et al. \ 2002, A\&A, 384, 383 

\bibitem[Jaffe \& Perola(1973)]{1973A&A....26..423J} 
Jaffe, W.~J., \& Perola, G.~C.\ 1973, A\&A, 26, 423 

\bibitem[Katz-Stone et al.(1993)]{1993ApJ...407..549K}
Katz-Stone, D.~M., Rudnick, L., \& Anderson, M.~C.\ 1993, ApJ, 407, 549 

\bibitem[Ogrean et al.(2013)]{2013MNRAS.429.2617O} 
Ogrean, G.~A., Br{\"u}ggen, M., R{\"o}ttgering, H., et al.\ 2013, MNRAS, 429, 2617 

\bibitem[{Pacholczyk}(1970)]{pacholcyzk}
{Pacholczyk}, {A.~G}, 1970, Radio astrophysics 

\bibitem[Rees(1989)]{1989MNRAS.239P...1R}
Rees, M.~J.\ 1989, MNRAS, 239, 1P 

\bibitem[{Shioya} {et~al.}(2008)]{2008ApJS..175..128S}
{Shioya}, {Y.}, {Taniguchi}, {Y.}, {Sasaki}, {S.~S.}, {et~al.} 2008, ApJS, 175, 128 

\bibitem[Stroe et al.(2013)]{2013A&A...555A.110S}
Stroe, A., van Weeren, R.~J., Intema, H.~T., et al.\ 2013, A\&A, 555, A110 

\bibitem[Stroe et al.(2014a)]{2014MNRAS.438.1377S}
Stroe, A., Sobral, D., R{\"o}ttgering, H., \& van Weeren, R.\ 2014a, MNRAS, 438, 1377 

\bibitem[Stroe et al.(2014b)]{AMI}
Stroe, A., Rumsey, C., Harwood, J. J., et al.\ 2014b, MNRAS, L63 

\bibitem[Umeda et al.(2004)]{2004ApJ...601..805U}
Umeda, K., Yagi, M., Yamada, S.~F., et al.\ 2004, ApJ, 601, 805 

\bibitem[{van~Weeren} {et~al.}(2010)]{2010Sci...330..347V}
{van~Weeren}, {R. J.}, {R{\"o}ttgering}, {H.}, {Br{\"u}ggen}, {M.}, \& {Hoeft}, {M.} 2010, Science, 330, 347
\end{thebibliography}

\label{lastpage}
\end{document}